\input psfig.sty
\documentstyle[12pt,twoside,fleqn,espcrc1,epsf]{article}

\newcommand{\AmS}{{\protect\the\textfont2
  A\kern-.1667em\lower.5ex\hbox{M}\kern-.125emS}}

\title{Four Fermion Models at Non-Zero Density}

\author{Simon Hands\address{Department of Physics, University of Wales
Swansea,\\
Singleton Park, Swansea SA2 8PP, U.K.}
\thanks{supported by EU TMR contract no. ERBFMRXCT97-0122}}

\begin{document}
\maketitle

\begin{abstract}
I review the properties of the three-dimensional Gross-Neveu model 
formulated with non-zero chemical potential and temperature, 
focussing on results 
obtained by lattice Monte Carlo simulation.
\end{abstract}

\section{INTRODUCTION}

This talk concerns simulations of fermionic systems at non-zero density,
which don't suffer from the problems associated with gauge theories. 
To paraphrase the sub-title of the workshop, they could be described as
``not-quite-so complex systems with a {\sl real\/} action''; however, 
as I propose to show, the physics involved is still rich.
The two theories I will discuss are referred to as ``Gross-Neveu'' models
\cite{GN},
and are described by the following Lagrangian densities:
\begin{eqnarray}
{\cal L}_A &=& \bar\psi_i({\partial\!\!\!/}\,+m)\psi_i-
{g^2\over{2N_f}}(\bar\psi_i\psi_i)^2 \\
{\cal L}_B &=& \bar\psi_i({\partial\!\!\!/}\,+m)\psi_i-
{g^2\over{2N_f}}\left[(\bar\psi_i\psi_i)^2-
(\bar\psi_i\gamma_5\psi_i)^2\right]
\end{eqnarray}
In each case the index $i$ runs over $N_f$ flavors of fermion, 
assumed to be four-component Dirac spinors. In the chiral limit $m\to0$,
symmetries akin to chiral symmetries can be identified:
\begin{equation}
A\;\;{\mbox Z}_2:\;\;\;\psi_i\mapsto\gamma_5\psi_i\;\;;\;\;
\bar\psi_i\mapsto-\bar\psi_i\gamma_5 \\
\label{eq:chZ2}
\end{equation}
\begin{equation}
B\;\;{\mbox U(1)}:\;\;\;\psi_i\mapsto\exp(i\alpha\gamma_5)\psi_i\;\;;\;\;
\bar\psi_i\mapsto\bar\psi_i\exp(i\alpha\gamma_5) 
\label{eq:chU(1)}
\end{equation}
A version of the model with a global $\mbox{SU(2)}_L\otimes\mbox{SU(2)}_R$ 
symmetry can also be
formulated.

I will focus on the case where the models (1,2) are formulated in three 
spacetime
dimensions, where the following features hold:

\begin{enumerate}

\item[$\bullet$] For sufficiently strong coupling $g^2$ the models exhibit
dynamical chiral symmetry breaking at zero temperature and density.

\item[$\bullet$] The spectrum of excitations contains both baryons and mesons, 
ie. the elementary fermions, and composite fermion -- anti-fermion states.

\item[$\bullet$] The models have an interacting continuum limit.

\item[$\bullet$] For model $B$, whose chiral symmetry is continuous, the 
spectrum contains Goldstone modes.

\item[$\bullet$] When formulated on a lattice, the models have real Euclidean
actions even for chemical potential $\mu\not=0$, and hence can be simulated 
by standard Monte Carlo techniques.
\end{enumerate}

For all these reasons, the models (1,2) are useful toys for understanding 
the behaviour of strongly-interacting matter at high density, as I shall
discuss.

\section{MEAN FIELD ANALYSIS}

It is considerably easier to treat the Gross-Neveu model, both analytically and
numerically, if the Lagrangians (1,2) are bosonised by introducing auxiliary
fields:
\begin{eqnarray}
{\cal L}^\prime_A &=& \bar\psi_i({\partial\!\!\!/}\,+m+\sigma)\psi_i+
{N_f\over{2g^2}}\sigma^2 \nonumber \\
{\cal L}^\prime_B &=& \bar\psi_i({\partial\!\!\!/}\,+m+\sigma+i\pi\gamma_5)
\psi_i+
{N_f\over{2g^2}}(\sigma^2+\pi^2) 
\label{eq:Lbos}
\end{eqnarray}
Since the scalar field $\sigma$ and pseudoscalar $\pi$ only appear quadratically
without derivative terms, it is trivial to integrate them out and recover an 
identical generating function to that derived from the original Lagrangian.
In the chiral limit $m=0$ the fermion may still have a dynamical mass
$\Sigma\equiv\langle\sigma\rangle$. We can solve for the expectation value
$\Sigma$ self-consistently via the {\sl Gap Equation\/}, given by the diagram
of Fig. 1a:
\begin{equation}
\Sigma=g^2\int_p\mbox{tr}{1\over{ip{\!\!\!/}\,+\Sigma}}
\label{eq:gap}
\end{equation}
We find a non-trivial solution $\Sigma\not=0$, which breaks the chiral
symmetries (3,4), if
\begin{equation}
{1\over g^2}<{1\over g_c^2}={{2\Lambda}\over\pi^2},
\end{equation}
where $\Lambda$ is the UV cutoff on the momentum integral in (\ref{eq:gap}).
Note that $\Sigma\to0$ as $g^2\nearrow g^2_c$.
This result is exact up to corrections of $O(1/N_f)$, and it turns out that
the self-consistent, or mean-field, approach coincides with the leading order of
a $1/N_f$ expansion.

\begin{figure}[htb]
\psfig{figure=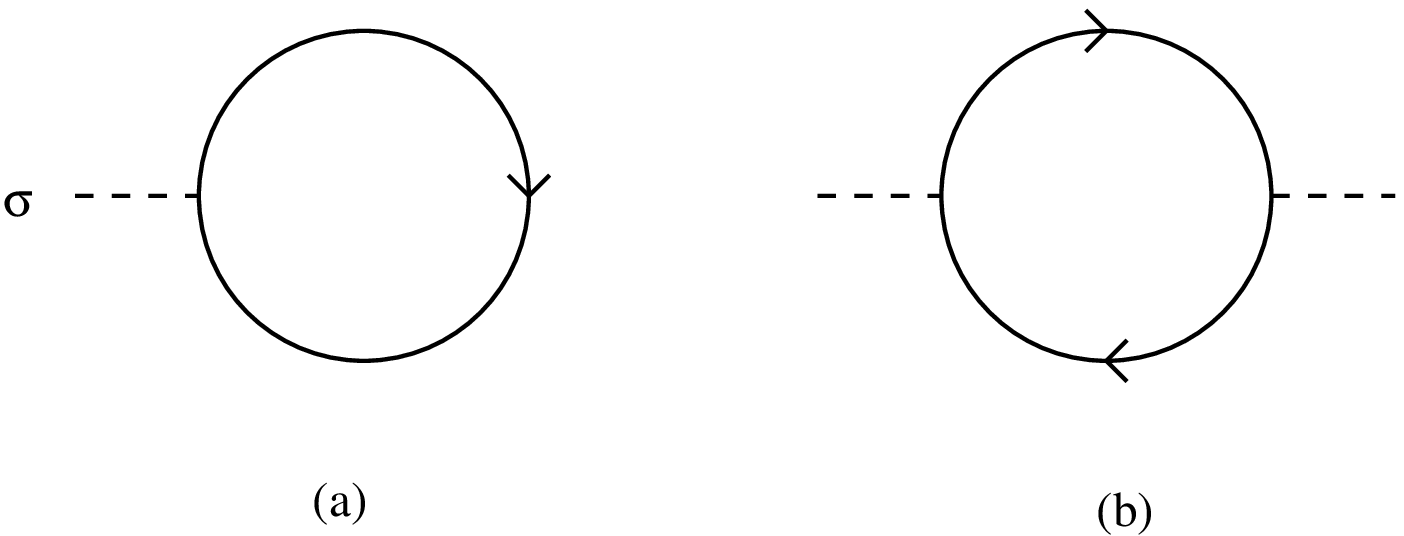,width=4.0in}
\vskip -10pt
\caption{Leading order diagrams in the GN model}
\label{fig:GNL}
\bigskip
\bigskip
\psfig{figure=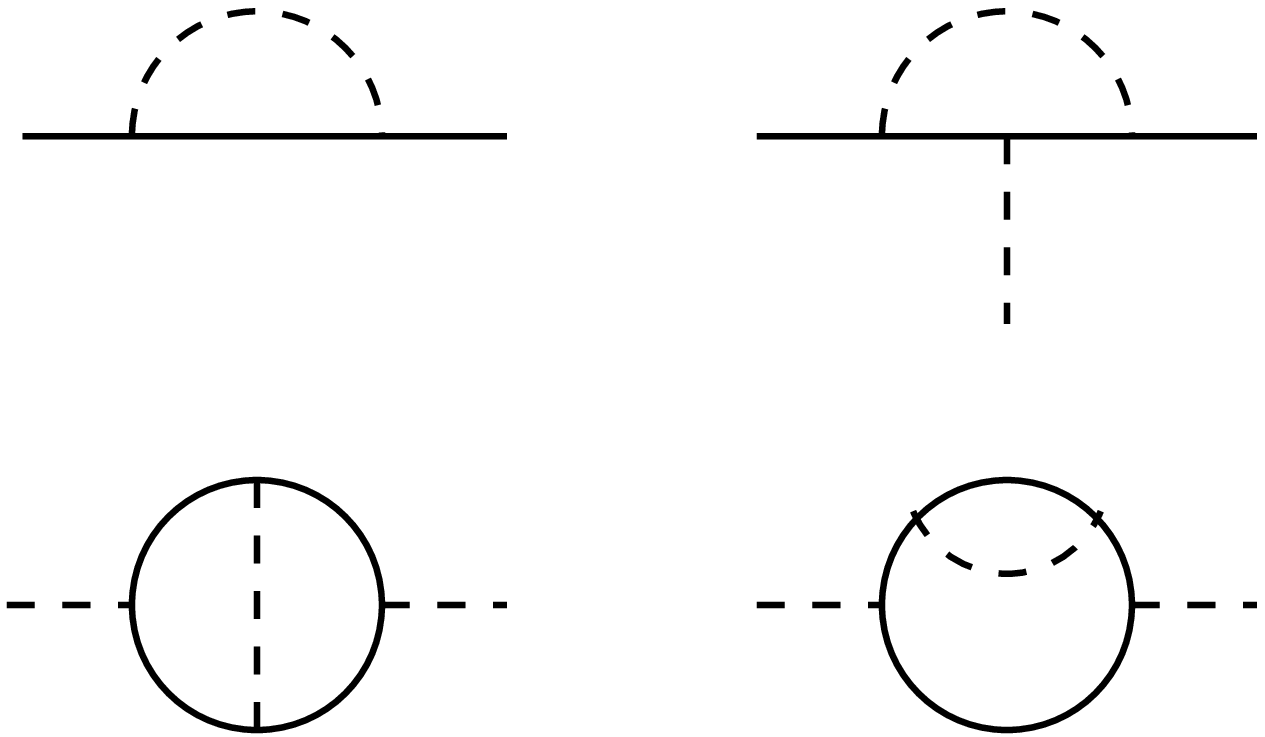,width=4.0in}
\vskip -10pt
\caption{$O(1/N_f)$ radiative corrections}
\label{fig:GNR}
\end{figure}

To the same leading order in $1/N_f$ there is a correction to the scalar
propagator (equal to $g^2/N_f$ at tree level) 
from the bubble diagram of Fig. 1b. Remarkably, the linear divergence in this
diagram is cancelled by the divergence in the definition of $1/g_c^2$, leading
to a closed-form expression which is finite 
when expressed in terms of $\Sigma$:
\begin{equation}
D_\sigma(k)={1\over N_f}{{2\pi\surd k^2}\over{(k^2+4\Sigma^2)
\tan^{-1}\left({\surd k^2}\over{2\Sigma}\right)}}.
\label{eq:D}
\end{equation}
For the $\pi$ field of model $B$, the factor $(k^2+4\Sigma^2)$ is replaced
by $k^2$. We can examine the behaviour of (\ref{eq:D}) in two limits.
In the IR regime $k\ll\Sigma$, we find
\begin{equation}
D_\sigma(k)\propto{1\over{k^2+(2\Sigma)^2}}\;\;\;;\;\;\;
D_\pi(k)\propto{1\over k^2},
\end{equation}
ie. the scalar resembles a weakly-bound fermion -- anti-fermion composite
whilst the pseudoscalar is a Goldstone mode. In the opposite UV limit
$k\gg\Sigma$, we have
\begin{equation}
D_{\sigma,\pi}(k)\propto{1\over{\surd k^2}}.
\end{equation}
Thus the UV behaviour of the propagators is {\sl softer\/} than would be
expected for an auxiliary field. Therefore higher order
diagrams, such as those of Fig. \ref{fig:GNR}, are less divergent
than might be expected by naive power-counting: the conseqence is that 
the divergences can all be absorbed by rescaling the parameters of the original
Lagrangian (\ref{eq:Lbos}), and the $1/N_f$ expansion is exactly renormalisable
\cite{RWP}.

The transition between chirally symmetric and broken phases at $g^2=g_c^2$
defines an ultra-violet fixed point of the renormalisation group. It is
characterised by non-Gaussian values for the critical exponents:
\begin{equation}
\beta=\nu=\gamma=\eta=1\;\;\;;\;\;\;\delta=2.
\label{eq:exp}
\end{equation}
Corrections to these values are $O(1/N_f)$ and calculable
\cite{HHK1}\cite{HHK1*}.
Indeed, they are currently known to $O(1/N_f^2)$ \cite{Gracey}, and 
when extrapolated to small values of $N_f$ are 
supported by Monte Carlo estimates
\cite{Biel}. The continuum limit $g^2\to g_c^2$ may be taken in either phase;
since our interest is in the restoration of the dynamically broken symmetry at
non-zero density, we shall always work with a value of $g^2$ corresponding
to the broken phase $\Sigma\not=0$, $\langle\bar\psi\psi\rangle\not=0$. The 
rationale will be to refer all calculated quantities to a mass scale given 
by the physical fermion mass; to leading order in $1/N_f$ this scale is
precisely $\Sigma$ \cite{HHK1*}.

The mean-field gap equation (\ref{eq:gap}) is readily generalised to non-zero 
temperature $T$ and chemical potential $\mu$: 
\begin{equation}
{1\over g^2}=4T\sum_{n=-\infty}^\infty\int{{d^2p}\over{(2\pi)^2}}
{1\over{((2n-1)\pi T-i\mu)^2+p^2+\Sigma^2(\mu,T)}}.
\end{equation}
The equation can be solved \cite{RWP2}, 
eliminating the coupling $g^2$ which sets the
cutoff scale in favour of the physical scale $\Sigma_0$, 
the value of $\Sigma$ at $\mu=T=0$:
\begin{equation}
\Sigma-\Sigma_0=-T\left[\ln(1+\mbox{e}^{-{{(\Sigma-\mu)}\over T}})
                       +\ln(1+\mbox{e}^{-{{(\Sigma+\mu)}\over T}})\right].
\label{eq:mft}
\end{equation}
\begin{figure}
\psfig{figure=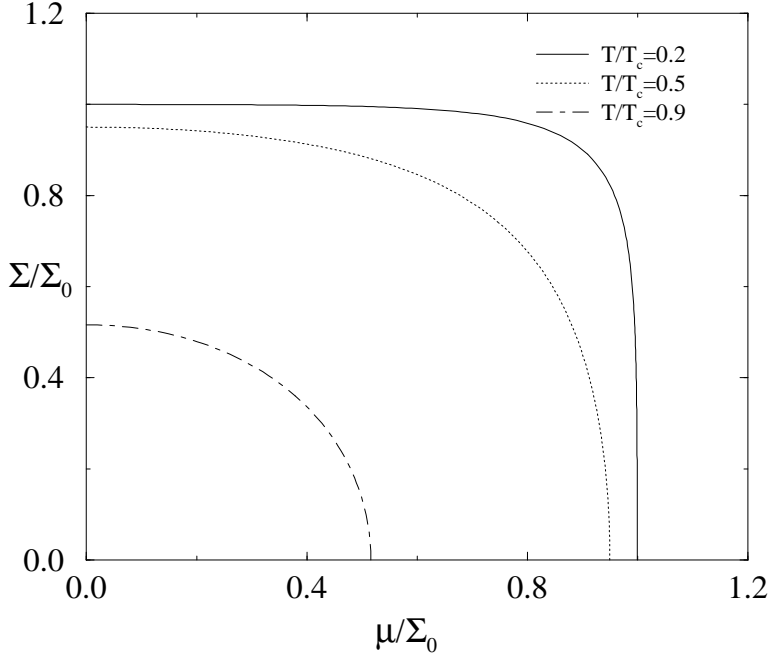,width=4.0in}
\vskip -20pt
\caption{Mean field solution for $\Sigma(\mu,T)$}
\label{fig:mft}
\end{figure}
The phase diagram of the model in the $(\mu,T)$ plane consists of a
chirally broken phase separated from a symmetric phase by the critical line
$\Sigma(\mu,T)=0$, ie:
\begin{equation}
1-{\mu\over\Sigma_0}=2{T\over\Sigma_0}\ln(1+\mbox{e}^{-{\mu\over T}}).
\label{eq:phase}
\end{equation}
For zero chemical potential we thus identify a critical temperature
\begin{equation}
T_c={\Sigma_0\over{2\ln2}},
\end{equation}
whereas for zero temperature we identify the behaviour
\begin{equation}
\Sigma(\mu)=\Sigma_0\theta(\mu_c-\mu)\;\;\;\mbox{with}\;\;\;
\mu_c=\Sigma_0.
\end{equation}
The phase transition is everywhere second order 
except at the isolated point $T=0$: this can be traced to the slope of 
the surface $\Sigma(\mu,T)$ diverging in an essentially singular way as 
$T\to0$, $\mu\to\mu_c$.
Solutions for $\Sigma(\mu)$ for various values of $T/T_c$ are shown in Fig.
\ref{fig:mft}.

\section{LATTICE SIMULATIONS FOR $\mu\not=0$}

The first Monte Carlo simulation of a four-fermi model with $\mu\not=0$
was for the two dimensional Gross-Neveu model \cite{Karsch}. The method is
easily extended to three dimensions, and I will survey results from
both models $A$ \cite{HKK2} and $B$ \cite{HKK3}. In all cases the staggered
fermion formulation, involving single-spin component
Grassmann fields $\chi,\bar\chi$ is used; due to species doubling we have the
result that in three dimensions $N$ flavors of staggered fermion correspond to
$N_f=2N$ continuum flavors \cite{BB}.
The free fermion kinetic term in the presence of
a chemical potential is $\bar\chi{\cal M}\chi$, with
\begin{equation}
{\cal M}^{ij}_{xy}={{\delta^{ij}}\over2}
\left[(\mbox{e}^\mu\delta_{yx+\hat0}-\mbox{e}^{-\mu}\delta_{yx-\hat0})
+\sum_{\nu=1,2}\eta_\nu(x)(\delta_{yx+\hat\nu}-\delta_{yx-\hat\nu})
+2m\delta_{xy}\right],
\end{equation}
and the sign factor $\eta_\nu(x)\equiv(-1)^{x_0+\cdots+x_{\nu-1}}$.

The lattice version of the interacting theory starts from the bosonised form of
the action (\ref{eq:Lbos}).
For technical reasons it is preferable to formulate the auxiliary scalar
fields $\sigma$ and $\pi$ on the dual lattice sites $\tilde x$ \cite{CER}.
For model $A$, the full fermion matrix including the interaction with the
scalars is thus
\begin{equation}
M^{ij}_{xy}={\cal M}^{ij}_{xy}+{1\over8}\delta_{xy}\delta^{ij}
\sum_{<\tilde x,x>}\sigma(\tilde x),
\end{equation}
where $<\tilde x,x>$ denotes the set of dual sites neighbouring 
$x$.  Note that $M$ is real and diagonal in the flavor
index $i$ running from 1 to $N$. Noting the chiral symmetry analogous
to (\ref{eq:chZ2}), valid for bare mass $m=0$:
\begin{equation}
\chi(x)\mapsto\varepsilon(x)\chi(x)\;\;\;;\;\;\;
\bar\chi(x)\mapsto\varepsilon(x)\bar\chi(x)\;\;\;\mbox{with}
\;\;\;\varepsilon(x)=
(-1)^{x_0+x_1+x_2},
\end{equation}
we see that the full global symmetry of the lattice action is
$\mbox{O}(N)\otimes Z_2$. In the continuum limit, we expect this symmetry 
to enlarge to $\mbox{U}(2N)\otimes Z_2$.
Now, the hybrid Monte Carlo algorithm
\cite{Duane} simulates fermions by integrating over bosonic ``pseudofermion''
fields, using a real bilinear action $\phi(M^\dagger M)^{-1}\phi$. Since
$M$ is real, the pseudofermion fields can be taken to be real, and the resulting
expression for the path integral on integrating over $\phi$ is
\begin{equation}
Z=\int\prod_{\tilde x}d\sigma(\tilde x)\sqrt{\mbox{det}M^{tr}M}\exp\left(
-{N\over{2g^2}}\sum_{\tilde x}\sigma^2(\tilde x)\right).
\label{eq:Z}
\end{equation}
For $N$ even, the path integral measure due to the fermions is thus
precisely $\mbox{det}M$ as required. 

For model $B$ the discussion is slightly more subtle, because now the matrix
$M$ is complex:
\begin{equation}
M^{ij}_{xy}={\cal M}^{ij}_{xy}+{1\over8}\delta_{xy}\delta^{ij}
\sum_{<\tilde x,x>}\left(\sigma(\tilde x)+i\varepsilon(x)\pi(\tilde x)
\right).
\label{eq:latu1}
\end{equation}
The weight in the functional integral after integrating over complex
pseudofermion fields is thus
$\mbox{det}({\cal M}+\sigma+i\varepsilon\pi)
\,\mbox{det}({\cal M}+\sigma-i\varepsilon\pi)$, and is explicitly real 
(because the only complex terms occur on the diagonal of $M$). However, the
simulation describes $N$ flavors of fermion with positive axial charge and
$N$ flavors with negative axial charge, and the global symmetry including
(\ref{eq:chU(1)}) is $\mbox{U}(N)\otimes\mbox{U}(N)\otimes\mbox{U}(1)$.
In the continuum limit this should enlarge to
$\mbox{U}(2N)\otimes\mbox{U}(2N)\otimes\mbox{U}(1)$, but it is important
to note that the $\mbox{U}(N_f)\otimes\mbox{U}(1)$ symmetry of (2)
is {\sl not} recovered. Even for $\mu=0$, the action (2) leads to a complex
measure, and cannot be simulated by this algorithm.

\begin{figure}
\psfig{figure=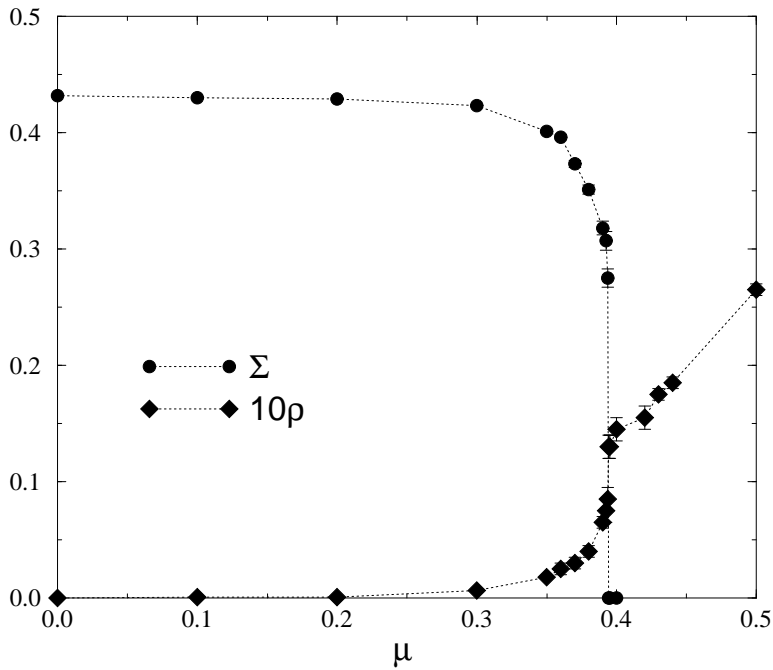,width=4.0in}
\vskip -20pt
\caption{$\Sigma$ vs. $\mu$ for $1/g^2=0.7$}
\label{fig:sig_mu}
\end{figure}

Model $A$ has been simulated using the partition function 
(\ref{eq:Z}) with $N_f=12$
\cite{HKK2}. Since the chiral symmetry group is discrete, and since the 
diagonal elements of $M$ are in general non-zero, it is possible to work 
directly in the chiral limit $m=0$. Fig. {\ref{fig:sig_mu} 
shows $\Sigma(\mu)$ evaluated
on a $20^3$ lattice (approximating $T=0$) with coupling $1/g^2=0.7$,
corresponding to the system being in the broken phase at zero density 
with $\Sigma_0=0.43$ (the bulk critical coupling $1/g_c^2$ is estimated 
to be 0.98(3)). We see that the order parameter
$\Sigma$ is very weakly dependent on $\mu$
up to a critical value $\mu_c\simeq0.394\simeq\Sigma_0$, whereupon it drops to
zero dramatically, suggesting a strong first order transition in 
accordance with the mean-field prediction. Also plotted is
the fermion number density $\rho$, defined by 
\begin{equation}
\rho(\mu)=-{1\over V}{{\partial\ln Z}\over{\partial\mu}}=
{1\over{2V}}\biggl\langle\sum_x\left(\mbox{e}^\mu M^{-1}_{x,x+\hat0}+
            \mbox{e}^{-\mu}M^{-1}_{x,x-\hat0}\right)\biggr\rangle,
\end{equation}
which shows a sharp rise across the symmetry-restoring transition.
It is tempting to interpret the $\rho\not=0$ signal for $\mu<\mu_c$ as
evidence for a ``nucleon liquid'', but this would be premature, since it is
quite possibly a finite volume (ie. non-zero temperature) artifact. Indeed,
test simulations on a $12^3$ system show that the transition is 
softened considerably on a smaller volume, with $\rho$ increased for
$\mu<\mu_c$.
It is also possible to study the approach to the continuum limit. Since
$\mu$ couples to a conserved current, it should not undergo renormalisation,
and hence $\mu_c$ should scale like a physical mass, leading to the relation
\begin{equation}
\mu_c\propto\left({1\over g_c^2}-{1\over g^2}\right)^\nu,
\label{eq:nu}
\end{equation}
with $\nu$ the correlation length critical exponent. From runs at 
$1/g^2=0.7,0.75,0.8$ we find relation (\ref{eq:nu}) verified with
$\nu=1.05(10)$, in agreement with the leading order prediction (\ref{eq:exp}).

\begin{figure}
\psfig{figure=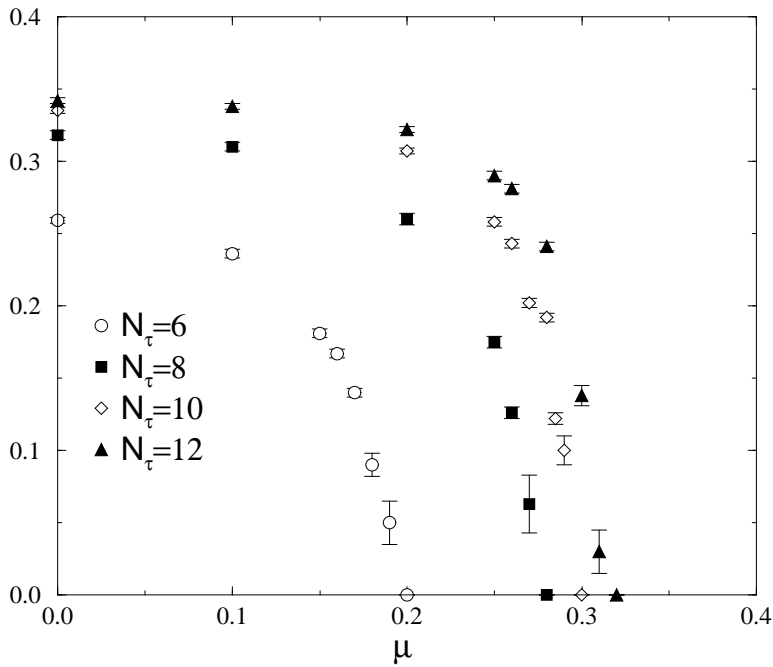,width=4.0in}
\vskip -20pt
\caption{$\Sigma$ vs. $\mu$ on asymmetric lattices}
\label{fig:sig_muT}
\end{figure}
\begin{figure}
\psfig{figure=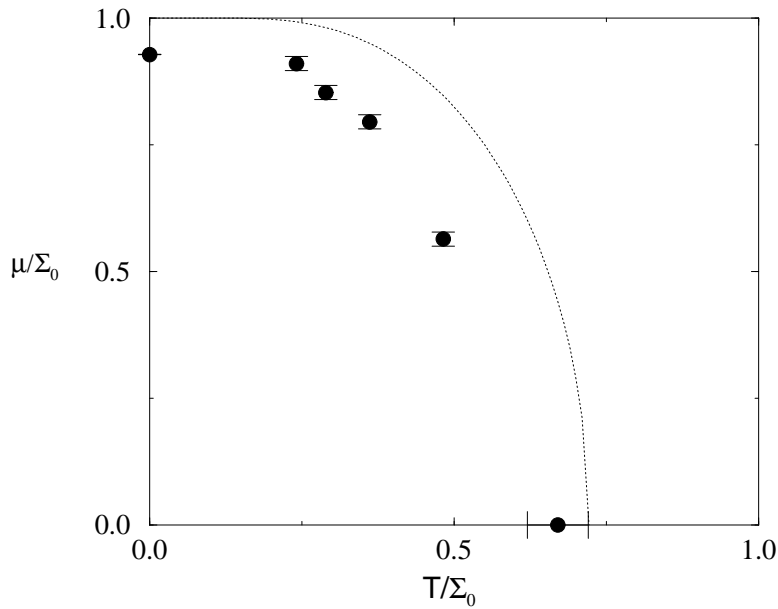,width=4.0in}
\vskip -20pt
\caption{Phase diagram in the $(\mu,T)$ plane}
\label{fig:phase}
\end{figure}

We have also studied the transition at non-zero temperature by simulations
on asymmetric lattices $N_\tau\times36^2$ with $1/g^2=0.75$. The results are 
shown in Fig.~\ref{fig:sig_muT}. The data show that the transition is softened,
with $\mu_c$ a decreasing function of temperature, in qualitative agreement
with the mean field prediction. Estimates of the location of the phase
transition are plotted in Fig \ref{fig:phase} together with the mean field
prediction (\ref{eq:phase}), showing reasonable agreement; the discrepancies
can probably be accounted for by $O(1/N_f)$ corrections.

There are two issues raised by the lattice simulations that are worth
discussing. First, what is the order of the transition for $T>0$? The isolated
nature of the first order 
point predicted by mean field theory seems unlikely to persist once
quantum effects are included in the calculation. Indeed, the simulations
showing a two-state signal characteristic of the first order transition
were performed on lattices of finite extent, and by implication at a small but
non-zero temperature. The most likely 
scenario is for the line of first order transitions to extend into the
$(\mu,T)$ plane, to end at 
a tricritical point separating it from a line of second order transitions
continuing to the point $(0,T_c)$.
This behaviour has been verified by
calculation of $O(1/N_f)$ corrections in the two dimensional model
\cite{Wolff}, although ironically, the thermodynamics in this case are not
described by the $1/N_f$ expansion due to condensation of topological
defects in the resulting one dimensional system \cite{Raj}\cite{Karsch}.
Numerical studies with higher precision, or explicit
calculations of $O(1/N_f)$
corrections in three 
dimensions would be useful to clarify the situation further.

Secondly, what is the nature of the pure thermal phase transition, ie. along the
$\mu=0$ axis? The mean field solution (\ref{eq:mft}) predicts a transition
with Gaussian exponents $\beta=\nu={1\over2}$, $\delta=3$, $\gamma=1$, $\eta=0$,
and indeed, early simulations \cite{HHK1*}
on a $12\times36^2$ system with $N_f=12$ and with $m\not=0$, estimated
$\delta=3.0(1)$ in apparent confirmation. This led to speculation 
\cite{KK} that models
in which a composite field (in this case a bound fermion -- anti-fermion pair)
condenses violate the conventional wisdom \cite{PW}, which
predicts that the thermal transition has the exponents of the dimensionally
reduced ferromagnetic model with the same global symmetry,
in this case the two dimensional Ising model.
More recently there has been an analytic study \cite{Reisz} using both 
dimensional reduction and subsequent linked cluster expansion techniques, 
and a more
accurate numerical study \cite{Misha}: both favour the $2d$
Ising model exponents
for $N_f<\infty$. 

\begin{figure}
\psfig{figure=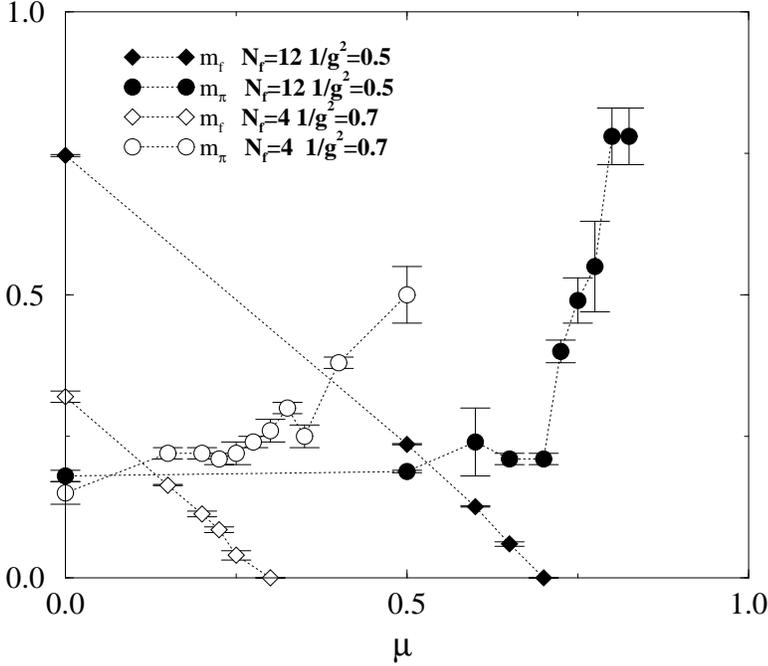,width=4.0in}
\vskip -20pt
\caption{Spectrum results for the U(1) model}
\label{fig:spectrum}
\end{figure}
Next I turn to numerical studies of Model $B$ \cite{HKK3}, 
which has a continuous U(1)
chiral symmetry and hence a Goldstone pion in the broken phase.
Here all simulations were performed on a $16^3$ system with bare mass $m=0.01$,
but this time with differing numbers of flavors, ie. $N_f=4$ and 12 in continuum
normalisation. 
Just as for the $\mbox{Z}_2$ model, it was found that a sharp chiral
symmetry restoring transition occurs at a critical $\mu_c\sim O(\Sigma_0)$, 
and that bulk expectation values of number density $\rho$ and energy density
were in good qualitative agreement with mean field theory, the greater 
discrepancy occuring as expected for $N_f=4$. More interesting results
come from studies of
the spectrum; the fermion mass $m_f$ is measured by studying decay of
the timeslice propagtor $G_+(t)$ in the timelike direction 
(a $\pm$ sign denotes that $G$ is calculated with a factor
$\mbox{e}^{\pm\mu}$ on forward-pointing links), while the
decay of the auxiliary field correlator yields the 
pion mass $m_\pi$. Results are shown in Fig. \ref{fig:spectrum} for
$N_f=12$, $1/g^2=0.5$ (ie. small quantum corrections, large lattice spacing), 
and $N_f=4$, $1/g^2=0.7$ (larger quantum effects, closer to the continuum
limit).

What is observed is a clear separation between the scale defined by the
critical chemical potential $\mu_c\sim m_f(\mu=0)$, and the scale $m_\pi/2$.
This is significant, since it implies that the Gross-Neveu model does {\sl
not\/} suffer
from the early onset of symmetry restoration observed in simulations of
both quenched
\cite{IMB1} and full \cite{IMB2} QCD. Instead we find $m_f(\mu)\simeq
m_f(0)-\mu$, and $m_\pi$ roughly constant for $\mu<\mu_c$, and rising sharply
for $\mu>\mu_c$, in accordance with naive expectation.
Why should this be? The answer seems to be that the Goldstone mechanism in these
models is realised through a pole in the pseudoscalar channel formed from
disconnected diagrams (ie. sum of bubbles), 
in accord with the $1/N_f$ expansion. Indeed, 
the connected (ie. flavor non-singlet) diagram, 
formed from the expectation $\langle
G_+(t)G_-^\dagger(t)\rangle$, appears numerically much less important,
yielding a bound state mass of roughly $2m_f(0)$ \cite{MPL}.
If the connected diagram had been numerically dominant, as is the case in
gauge theories, then the same diagram would have predicted a light state 
formed between two 
fermions of opposite axial charge (recall the discussion below
eqn. (\ref{eq:latu1})); since this state carries net fermion number, it would
be equivalent to a {\sl baryonic pion\/}, and hence promote a premature
restoration of the symmetry as $\mu$ increases \cite{MPL2}.

\section{SUMMARY AND OUTLOOK}

The Gross-Neveu model presents an ``easy'' problem, in the sense that it 
has a well-defined continuum limit, and can be explored in a controlled
fashion 
without introducing 
ad hoc interactions or form factors. Nonetheless, the physics
it contains for $\mu\not=0$ is rich, and is possibly not too dissimilar, at
least qualitatively, from that of QCD. Issues such as the nature of the thermal
transition, the existence of a tricritical point, and also a possible nucleon
liquid-vapour phase transition (expected for $\mu<\mu_c$ for sufficiently
low temperatures) can all be addressed by more thorough simulations than those
reviewed here. Moreover, quantitative information is also available in
principle from the $1/N_f$ expansion, known to be accurate for $N_f$ as small as
2 \cite{Biel}. In addition, the model might also serve as a useful warm-up
exercise for other analytic methods such as the Schwinger-Dyson or exact
renormalisation group approaches. 
Despite its apparent easiness, however, the model also presents a non-trivial
test for numerical algorithms, such as the Glasgow reweighting method
\cite{IMB2}, designed to
treat systems with complex integration measures. One insight to come out of
the study \cite{MPL} is that the Gross-Neveu model appears to be simulable 
by conventional means
because the Goldstone mechanism is realised in a fundamentally
different way to the case in
gauge theories, reflecting in turn the fact that the global symmetries of the
two problems are different for general $N_f$.

Let me finish by advertising the fact that there is also a ``difficult'' problem
in three dimensional fermion physics, namely the Thirring model, with
the following Lagrangian density:
\begin{equation}
{\cal L}_{Thirring} = \bar\psi_i({\partial\!\!\!/}\,+m)\psi_i+
{g^2\over{2N_f}}(\bar\psi_i\gamma_\mu\psi_i)^2. 
\end{equation}
This model has been studied by $1/N_f$ expansion \cite{Gomes}, 
Schwinger-Dyson equations \cite{Itoh}, 
and numerical simulation \cite{DHM}. I believe it to be the
simplest fermionic system for which a numerical solution is necessary.
Its essential features are:

\begin{enumerate}

\item[$\bullet$] It is non-perturbative in $1/N_f$.

\item[$\bullet$] It exhibits dynamical chiral symmetry breaking and
a UV fixed point of the renormalisation group for $N_f\leq4$.

\item[$\bullet$] The chiral symmetry group is $\mbox{U}(N_f)$, similar
to QCD.

\item[$\bullet$] For $\mu\not=0$ the action is complex.

\end{enumerate}

In other words, the model exhibits all the essential physics
{\sl except} confinement. Perhaps it would provide a suitable
intermediate challenge while we wait for the means to solve the real problem.

\end{document}